\documentclass[fleqn,10pt]{wlscirep}
\usepackage[utf8]{inputenc}
\usepackage[T1]{fontenc}
\title{Packaging code for reproducible research in the public sector}

\author[1, 2,*,a]{Federico Botta}
\author[2, 3, 4, b,$\dagger$]{Robin Lovelace}
\author[5, $\dagger$]{Laura Gilbert}
\author[6, c,$\dagger$]{Arthur Turrell}
\affil[1]{Department of Computer Science, University of Exeter, Exeter, UK}
\affil[2]{The Alan Turing Institute, London, UK}
\affil[*]{f.botta@exeter.ac.uk}
\affil[a]{\url{https://orcid.org/0000-0002-5681-4535}}
\affil[3]{Institute of Transport Studies, University of Leeds, UK}
\affil[4]{Active Travel England, UK}
\affil[b]{\url{https://orcid.org/0000-0001-5679-6536}}
\affil[5]{10DS, 10 Downing Streeet, UK}
\affil[6]{Data Science Campus, Office for National Statistics, UK}
\affil[c]{\url{https://orcid.org/0000-0002-2525-0773}}

\affil[$\dagger$]{The views expressed are those of the authors and may not reflect the views of Active Travel England, 10DS, the Office for National Statistics and the wider UK government.}


\begin{abstract}
The effective and ethical use of data to inform decision-making offers huge value to the public sector, especially when delivered by transparent, reproducible, and robust data processing workflows. One way that governments are unlocking this value is through making their data publicly available, allowing more people and organisations to derive insights. However, open data is not enough in many cases: publicly available datasets need to be accessible in an analysis-ready form from popular data science tools, such as R and Python, for them to realise their full potential.
\\ This paper explores ways to maximise the impact of open data with reference to a case study of packaging code to facilitate reproducible analysis. We present the \texttt{jtstats} project, which consists of R and Python packages for importing, processing, and visualising large and complex datasets representing journey times, for many modes and purposes at multiple geographic levels, released by the UK Department of Transport. \texttt{jtstats} shows how domain specific packages can enable reproducible research within the public sector and beyond, saving duplicated effort and reducing the risks of errors from repeated analyses. We hope that the \texttt{jtstats} project inspires others, particularly those in the public sector, to add value to their data sets by making them more accessible.
\end{abstract}
\begin{document}

\flushbottom
\maketitle
%
%
\thispagestyle{empty}

\section{Introduction}
Recent years have seen an increasing volume of data being collected and generated, in a phenomenon that has been labelled a `data revolution' \cite{kitchin_data_2014}. Our interactions with socio-technological systems create a vast amount of data on our behaviour, actions, society, and other aspects of our lives \cite{conte2012manifesto, Lazer2009, Vespignani2009, Botta2021, bannister2021rapid}. Governments, statistical agencies, international organisations and non-profit entities also collect, produce and release large volumes of data on the state of society, nations, economies, and a whole suite of useful indicators. To analyse these data sets and deliver the maximum amount of insight from them increasingly requires knowledge of data science techniques \cite{lovelace2017propensity, lovelace2019geocomputation}. However, in order to unlock the full value of open data, it is becoming apparent that merely making it available is a necessary but not sufficient condition as the complex nature of many data sets requires them to be shared in formats that are easy to work with in the popular analytical tools used in data science.
\\\\ Here, we use a UK-based data set that is regularly released by the UK Department for Transport (DfT) as an example of how publicly available data can be coupled with publicly available code to maximise value--including to the public sector bodies who may be most interested in using it. It is important to note that there is nothing special about these data or this example--it merely serves to demonstrate a point, and we could have chosen any number of other examples of data released by other public sector bodies. Here, though, our case study is data released by DfT on journey time statistics. To show the additional insights that can be gained when data are more accessible, we compare these to data on relative deprivation and so provide a way of assessing the accessibility of places and services with socio-economic information of the population. Data on relative deprivation is retrieved directly from the UK Department for Levelling Up, Housing and Communities website.
\\\\ In our particular case study of journey time statistics, the data are crucial for understanding a broad range of questions in both academic research and policy. For example, the interplay between travel times and socio-economic inequalities is particularly acute in some types of area where inequalities may be exacerbated by poor access to education or services. Therefore, the availability and accessibility of data (and tools) to study these issues is crucial for improving our cities and rural areas. The open source software presented here aims to provide simple tools to perform this analysis and to make these valuable data more easily accessible for research. Starting from a relatively complex data structure, we provide open tools that allow most data scientists to work with these data using standard data science pipelines.

The data sets that we attempt to make more accessible in this paper, and the open source software that facilitates this within a reproducible research pipeline, are relevant to a wide range of academic research and policy questions.
Variable travel times underlie many social phenomena, ranging from spatial variability in energy use \cite{breheny_compact_1995} to inequalities in educational opportunities \cite{moreno-monroy_public_2017}.
This makes open data on journey times valuable, particularly when provided at high spatial and temporal resolution, and when provided with reference to a wide range of transport modes and purposes.
\\\\ A further added value of making data sets openly available is that they can be combined and enriched by other existing open data sets. To demonstrate this, the Python version of the package presented here is also able to retrieve data on relative deprivation level directly from the government website. The English Indices of Multiple Deprivation (IMD) is published on the Department for Levelling Up, Housing and Communities' (formerly the Ministry of Housing, Communities \& Local Government) website and provides a measure of relative deprivation of areas in England across seven different domains: Income; Employment; Education; Health; Crime; Barriers to housing \& services; and Living environment.
\section{Data -- Journey time statistics (JTS)}
This data set is released by DfT and contains statistics on modelled journeys to key services in England, such as employment centres, schools and hospitals. The data is broadly divided into three key statistics: average minimum journey time, which is the shortest travel time to a specific service by mode of transport (car, walking, cycling, or public transport); origin indicators, measuring the number of different services in an area that can be reached in a given time; and destination indicators, measuring the proportion of users that can access a service in a certain time. The data is provided at different levels of spatial aggregation, from Lower Layer Super Output Areas (LSOA) to national, regional and local authority level.
\\\\ It is important to highlight that the JTS data does not represent actual journeys, but rather idealised trips generated by DfT using commercial software. In particular, the journeys are modelled based on the average journey completed on a Tuesday in the second week of October of the corresponding year during the morning peak travel time between 7am and 10am. More information on the JTS data and how the journeys are calculated and aggregated can be found on the DfT website \cite{jts}.
\\\\ The data are released by DfT in the \textit{Open Document Format}, more specifically \textit{Open Document Spreadsheet}, which is an open file format for sharing documents, texts, and spreadsheets. However, the complex nature of this data set, and the specific formatting of some of the files contained in it, make it difficult for data scientists to easily import the data for analysis. The open software that we released alongside this paper first converts the files to the simpler \texttt{.csv} format, and then processes it to make it available in other standard formats used in data science analysis. As discussed in the Conclusion section below, future work should consider using state-of-the-art open formats, such as Apache's Parquet format, to improve performance and inter-operability.
\section{Code}
\begin{figure}
    \centering
    \includegraphics[width = 0.4\textwidth]{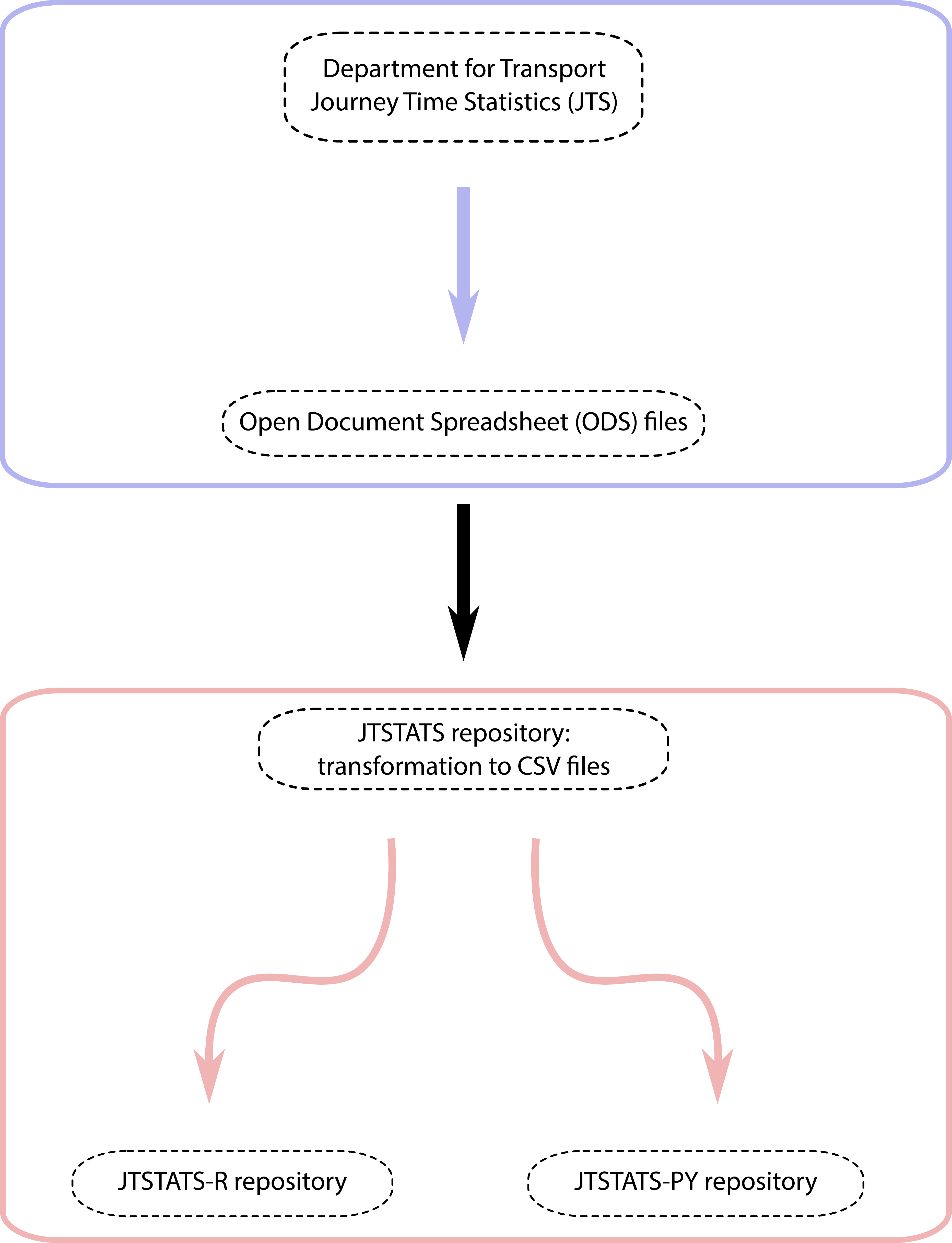}
    \caption{\textbf{Code and data workflow $\vert$} Packaging code and data to enable easy access to and analysis of government data is becoming increasingly important to maximise the value to public sector bodies of making their data openly available. Here, we use data on Journey Time Statistics (JTS) from the Department for Transport (DfT) as a case study. The data is generated by DfT and released publicly in \textit{Open Document Spreadsheet} format (green box at top). We then convert the data to a more user friendly \textit{csv} format, and use that as the basis of our \textit{R} and \textit{Python} packages (red box at bottom).}
    \label{fig:workflow}
\end{figure}

The open source software presented in this paper can be integrated into a reproducible research pipeline that can then be used by academics and policymakers alike. The software, both in its Python and R versions, is built in a simple, modular structure to allow ease of use as well as to enable data scientists to easily add features over time.
\\\\ Both implementations of the software rely on a version of the JTS data which has been converted from the original \texttt{.ods} file format to a more standard \texttt{.csv} format via a command line script (also openly available on the GitHub repository hosting the R and Python packages). Figure \ref{fig:workflow} depicts an overview of the workflow of the packages presented here.

\subsection{R}

\texttt{jstats} has been implemented as an R package which can be installed and loaded with the following commands from the R console (we recommend using a modern IDE such as RStudio or VSCode):

\texttt{
\\install.packages("remotes")
\\remotes::install\_github("datasciencecampus/
jtstats-r")
\\library(jtstats)
}
\\\\ After the package has been installed and loaded, you can start using it to get any of the 192 tables available from the JTS project.
You can get the table JTS0101 with the following command:
\\\\\texttt{
jts0101\_data = get\_jts(code = "JTS0101")
}
\\\\You can get geographic datasets for relevant tables by setting the geo argument to TRUE, as follows:
\\\\\texttt{
lsoa\_employment = get\_jts(type = "jts05",\\ purpose = "employment", sheet = 2017)
}
\begin{figure*}[t]
    \centering
    \includegraphics[width = 0.4\textwidth]{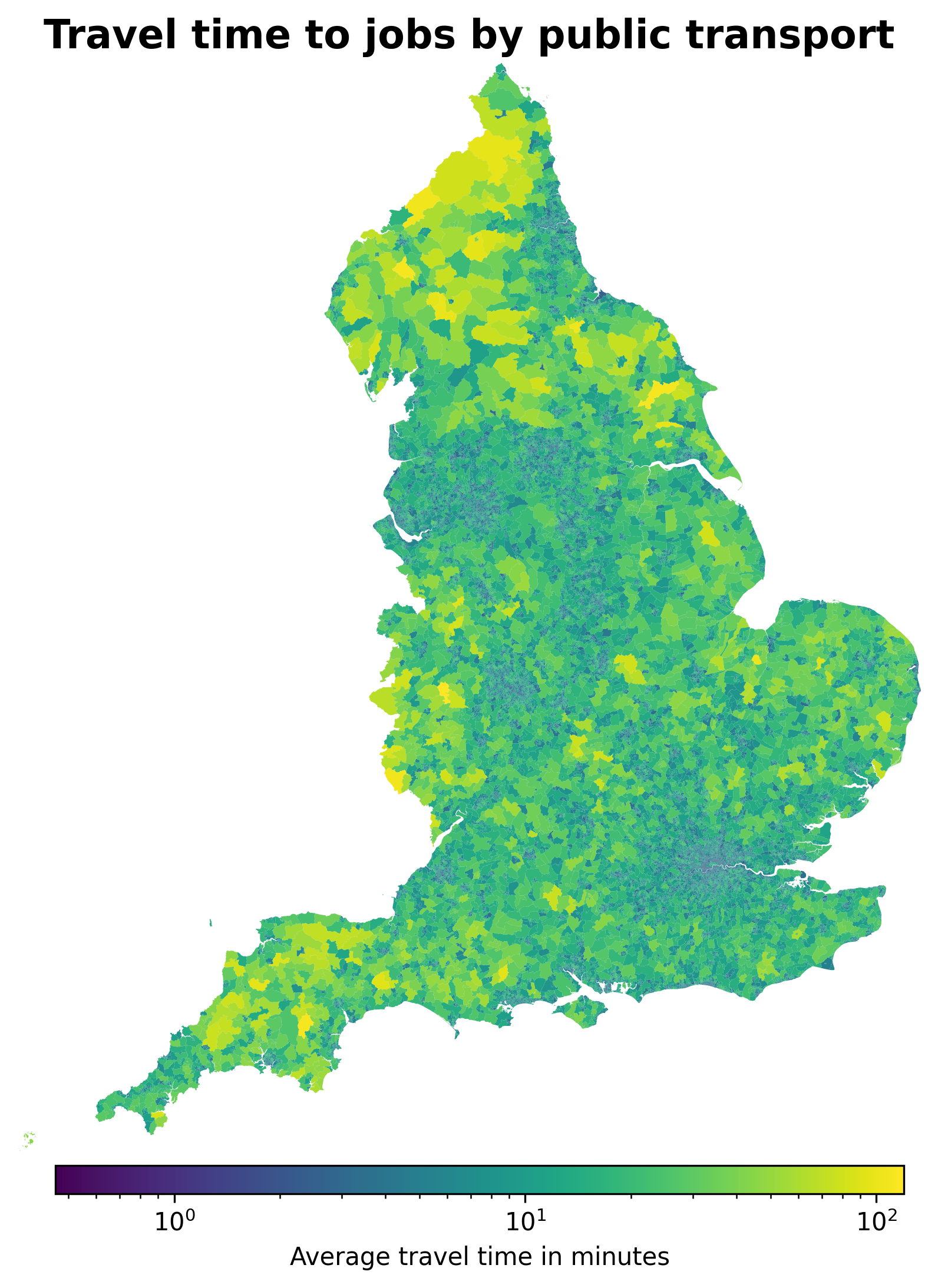}~
    \includegraphics[width = 0.4\textwidth]{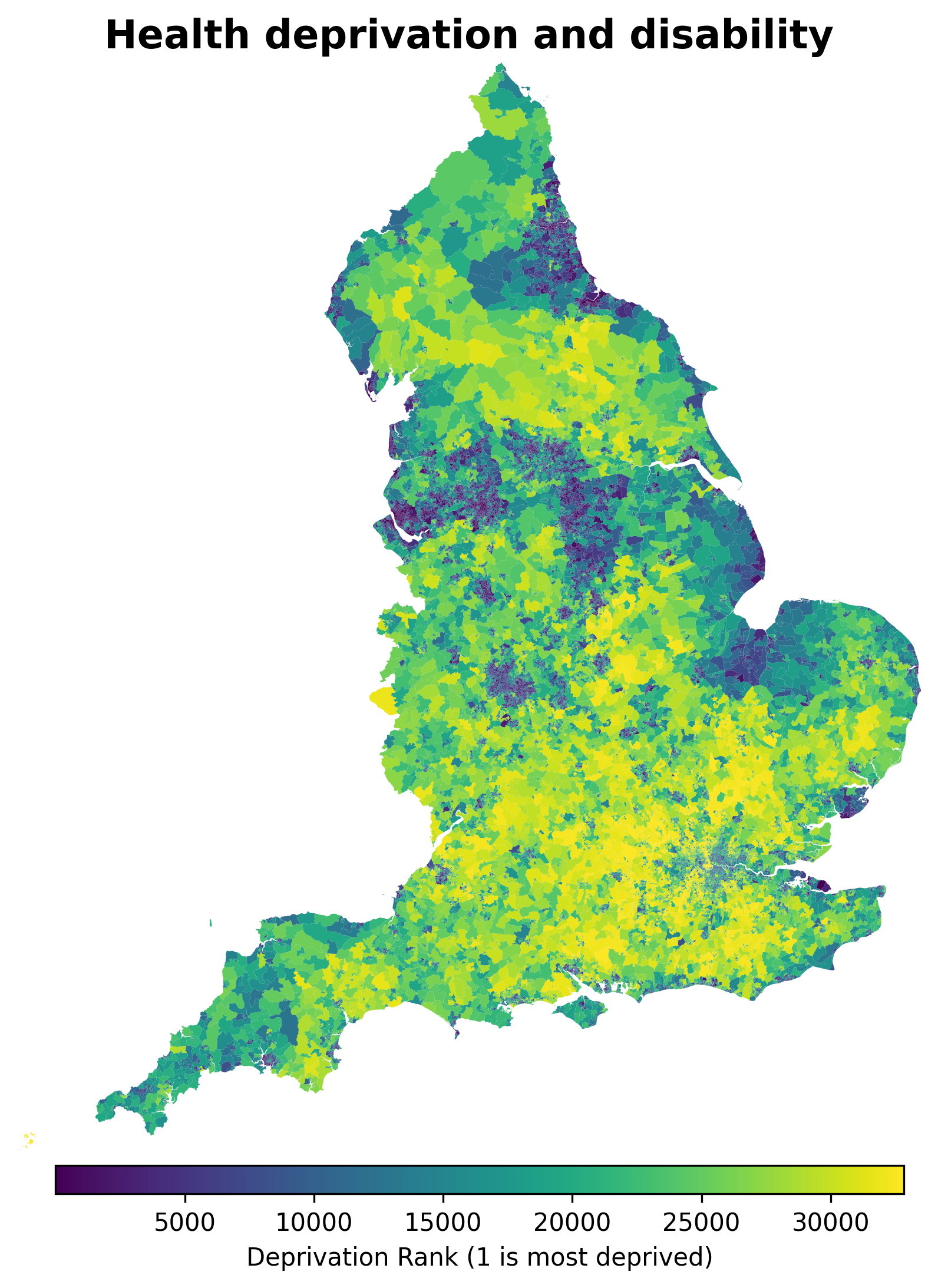}
    \caption{\textbf{Journey time statistics and deprivation data} We demonstrate the potential use of our package by depicting two variables from the data sets made easily available for analysis within our code. \textit{(left)} Here, we depict the travel time to employment centres with 100 to 499 jobs by public transport as made available by the \textit{Journey Time Statistics} data set published by the \textit{Department for Transport}. Travel time is capped to a maximum of 120 minutes. \textit{(right)} We also present data on relative deprivation. In particular, we show here deprivation ranks related to health and disability, as published by the Ministry of Housing, Communities \& Local Government in 2019.}
    \label{fig:maps_jts_imd}
\end{figure*}
\subsection{Python}
The Python implementation of this package has been developed in a Poetry environment running Python 3.9.1, and consists of four main parts:
\begin{itemize}
    \item \texttt{jts}: this implements the main functionalities to access the JTS data. It retrieves and cleans the required data based on the user input (see Table \ref{tab:python_table1} for an example), and returns it in a simple \texttt{pandas} data frame format, ready for analysis;
    \item \texttt{imd}: this retrieves the additional IMD data set based on the specified user input for the year and specific domain of deprivation;
    \item \texttt{geo}: this retrieves the spatial (GeoJSON) files for the LSOAs and local authorities;
    \item \texttt{plot}: this implements a few plotting functionalities, such as those used to generate the maps in Figure \ref{fig:maps_jts_imd}. This allows a simple, initial exploration of the data.
\end{itemize}

\begin{figure}[h!]
    \centering
    \includegraphics[width = 0.4\textwidth]{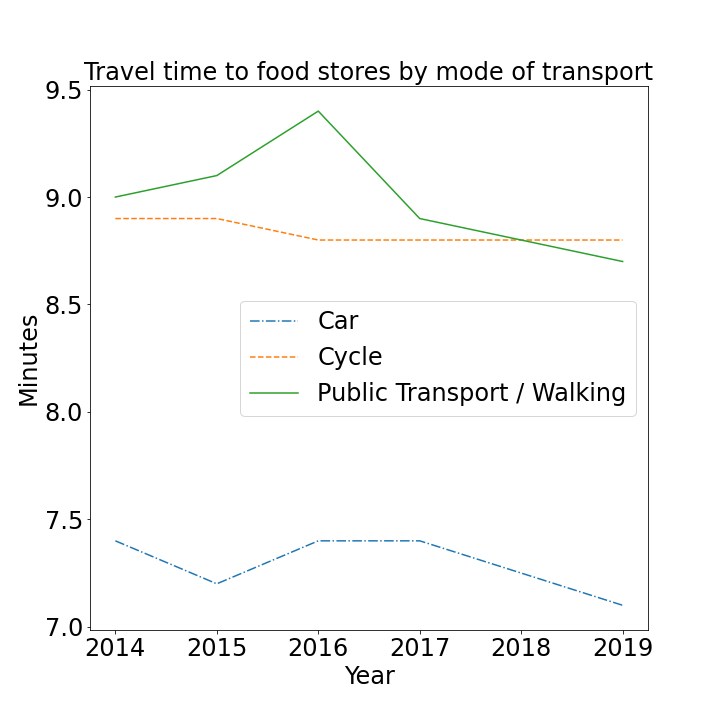}
    \caption{\textbf{Travel time to food stores by mode of transport $\vert$} Our package enables easy access to the data by returning a nicely formatted data frame which can be used for analysis. Here, we depict an example of the data which can be retrieved with our package: travel time to food stores over the years, disaggregated by mode of transport.}
    \label{fig:jts_year_food_mode}
\end{figure}

A brief tutorial script is also provided on the GitHub page to demonstrate basic functionalities and usage of this package (\url{https://github.com/datasciencecampus/jtstats-py}).
\\\\ Thanks to its modular structure, retrieving different data sets, or different variables within a data set, can be easily done within this software. For instance, the data needed for Figure \ref{fig:maps_jts_imd} can be retrieved by simply calling \texttt{jts.get\_jts(type\_code = 'jts05', purpose = 'employment', sheet = 2019)} , whereas the data for Figure \ref{fig:jts_year_food_mode} can be retrieved by \texttt{jts.get\_jts(table\_code = 'jts0101', sheet = 'JTS0101')}. Table \ref{tab:python_table1} provides a simple description of how the JTS05 tables can be retrieved using the Python module, along with some information on size of the retrieved data. Full information on how to retrieve the remaining JTS tables can be found at \url{https://github.com/datasciencecampus/jtstats}.
\\\\ To support further future development of the package, we also implemented, and provide alongside the package, a series of tests written using \texttt{pytest}. These test the retrieval of each JTS table by checking a pre-specified set of known values in the tables, so that any further development of the package can easily test any change implemented.
\\\\ A final note to mention regards missing values in some of the JTS tables. In particular, the JTS09 tables contains entries of the form `\texttt{..}', which have been replaced with 240 in the package; this is in agreement with the JTS data description that states that such entries correspond to trips of duration 240 minutes or more. Similarly, entries in the JTS0930 table which are missing (represented by `\texttt{--}' in the original files) have been replaced by \texttt{-99}.
\begin{table*}[h!]
\caption{\textbf{Retrieval of LSOA-level JTS data using the Python module} The Python version of the module allows easy retrieval of the JTS data using the \texttt{get\_jts()} function. The key parameters that can be specified in order to specify what data to retrieve are: \texttt{type\_code}, \texttt{spec}, \texttt{sheet} and \texttt{table\_code} (even though this last parameter is only rarely needed). The table here reports the values to be used for these parameters in order to retrieve the JTS data at the LSOA level, which are likely to be the most commonly used tables..The \texttt{sheet} parameter should be simply set to the year for which data is needed, e.g. \texttt{sheet = '2019'}. The \texttt{get\_jts()} function returns a \texttt{pandas} dataframe. The size of the returned dataframe is given in the last column (in the format \textit{\# rows $\times$\# columns}) of the table, for the specific case of 2019 data.}
\label{tab:python_table1}
\centering\small
\begin{tabular}{@{}llrrr@{}} \toprule 
Table title & \multicolumn{1}{c}{Description} & Type code & Spec/Sheet & Table size (\# rows $\times$\# columns) \\ \midrule
JTS0501 & Journey times for employment centres by mode of travel (LSOA) & jts05 & employment & 32844$\times$ 113 \\[5pt]
JTS0502 & Journey times for primary schools by mode of travel (LSOA) & jts05 & primary & 32844$\times$ 41\\[5pt]
JTS0503 & Journey times for secondary schools by mode of travel (LSOA) & jts05 & secondary & 32844$\times$ 41 \\[5pt]
JTS0504 & Journey times for further education by mode of travel (LSOA) & jts05 & further & 32844$\times$ 41 \\[5pt]
JTS0505 & Journey times for GPs by mode of travel (LSOA) & jts05 & gp & 32844$\times$ 41 \\[5pt]
JTS0506 & Journey times for hospitals by mode of travel (LSOA) & jts05 & hospital & 32844$\times$ 41 \\[5pt]
JTS0507 & Journey times for food stores by mode of travel (LSOA) & jts05 & food & 32844$\times$ 41 \\[5pt]
JTS0508 & Journey times for town centres by mode of travel (LSOA) & jts05 & town & 32844$\times$ 41\\[5pt]
JTS0509 & Journey times to pharmacy by cycle and car (LSOA) & jts05 & pharmacy & 32844$\times$ 23 \\[5pt]
\bottomrule
\end{tabular}
\end{table*}

\section{Conclusion}
Our experience of developing R and Python packages to enable the import of JTS data more easily demonstrates how software development can enable reproducible data science for public bodies and beyond. Government bodies are increasingly tasked with providing large and complex data sets to multiple stakeholders. 'Packaging-up' tools to enable direct access to them can support this task while building software development capacity and ensuring transparency and reproducibility of research with official datasets.

The JTS data files used here showcase the challenges faced by many public sector bodies wishing to obtain the maximum value from their data: developing documentation, providing pipelines for data cleaning, providing context (e.g. provision of geometries alongside entries representing administrative zones), and releasing data in the most relevant formats, all take extra resources. However, investing in these can enhance the ability of researchers and analysts to work with the data. The case study presented not only provides simple tools for accessing the data, but it also encourages a range of users by supporting two of the most popular languages for data science. 
The true potential of publicly available data sets can only be realised when a broad range of data scientists and analysts can work with them, with minimal barriers to entry. Other benefits of the approach include reducing risks of each researchers introducing errors into their work, leading to results that are not only erroneous, but hard for others to fix further down the `data pipeline'.

The approach we have taken is not without limitations.
Indeed, there are many additional improvements that could be made, including: the provision of intermediate data files in a compressed and analysis-ready format (such as Apache's Parquet format), enabling future proof and high performance in memory analysis using the multi-lingual Arrow framework (currently files are provided in .csv format); more rigorous tests to ensure that the outputs from R and Python implementations of \texttt{jstats} are the same; and better integration with upstream processes that led to the creation of the .ods files (raising the question of how to design data collection projects such as JTS around data science tools).
\\\\ The \texttt{jstats} project presented here is an example of how publicly available government data can be made available for reproducible analysis by developing open source tools packaged in popular and languages for data science. We urge others working with and publishing open datasets to develop packages for effective and transparent research. Publishing such packages alongside open datasets will add vast amounts of value to government digital assets that are in the public domain for the benefit of all.

\section*{Acknowledgements}
We would like to thank Greg Haigh (Advanced Analytics, Department for Transport) and Stephen Reynolds (Travel and Environment Data and Statistics (TRENDS) Division, Department for Transport) for feedback on the paper.
\\\\This work is funded by the Economic and Social Research Council (ESRC) \& ADR UK as part of the ESRC-ADR UK No.10 Data Science (10DS) fellowship in collaboration with 10DS and ONS (Federico Botta, grant number ES/W003937/1; Robin Lovelace, grant number ES/W004305/1)

\section*{Additional information}
The views expressed are those of the authors and may not reflect the views of Active Travel England, 10DS, the Office for National Statistics and the wider UK government.

\bibliography{references.bib}

\end{document}